\documentclass[prl,aps,twocolumn]{revtex4}

\usepackage{graphicx}
\usepackage{amsmath}
\usepackage{bm}
\usepackage{amstext}
\usepackage{amsxtra}

\usepackage{times}

\begin{document}

\title{Can a Bose gas be saturated?}

\author{Naaman Tammuz$^{1}$*, Robert P. Smith$^1$*, Robert L. D. Campbell$^1$, \\
Scott Beattie$^1$, Stuart Moulder$^1$, Jean Dalibard$^{1,2}$ \& Zoran Hadzibabic$^1$}

\affiliation{$^1$Cavendish Laboratory, University of Cambridge, J. J. Thomson Avenue, Cambridge CB3 0HE, U.K. \\
$^2$ Laboratoire Kastler Brossel, CNRS, UPMC, Ecole Normale Sup\'erieure, 24 rue Lhomond, F-75005 Paris, France}

\newcommand{\Nc}{N_{\rm c}}
\newcommand{\Ncid}{N_{\rm c}^{\rm (id)}}
\newcommand{\kB}{k_{\rm B}}
\newcommand{\Nt}{N_{\rm tot}}
\newcommand{\aho}{a_{\rm ho}}
\newcommand{\cs}{$\clubsuit$}
\newcommand{\Tcid}{T_{\rm c}^{\rm (id)}}

\date{\today}



\begin{abstract}
Bose--Einstein condensation is unique among phase transitions between different states of matter in the sense that it occurs even in the absence of interactions between particles. In Einstein's textbook picture of an ideal gas, purely statistical arguments set an upper bound on the number of particles occupying the excited states of the system, and condensation is driven by this saturation of the quantum vapour \cite{Einstein:1925a}. Dilute ultracold atomic gases are celebrated as a realisation of Bose--Einstein condensation in close to its purely statistical form \cite{Anderson:1995,Davis:1995c,brad97bec,Pethick:2002,Pitaevskii:2003}.  Here we scrutinise this point of view using an ultracold gas of potassium ($^{39}$K) atoms, in which the strength of interactions can be tuned via a Feshbach scattering resonance \cite{Roati:2007,Chin:2010}. We first show that under typical experimental conditions a partially condensed atomic gas  strongly deviates from the textbook concept of a saturated vapour. We then use  measurements at a range of interaction strengths and temperatures to extrapolate to the non-interacting limit, and prove that in this limit the behaviour of a Bose gas is consistent with the saturation picture. Finally, we provide evidence for the universality of our observations through additional measurements with a different atomic species, $^{87}$Rb. Our results suggest a new way of characterising condensation phenomena in different physical systems.
\end{abstract}

\maketitle

Einstein's prediction of a purely statistical phase transition follows directly from the Bose distribution for an ideal (non-interacting) gas of identical particles \cite{huan87,Pethick:2002,Pitaevskii:2003}. At a given temperature $T$, there is an upper bound $\Ncid(T)$ on the number $N'$ of bosons in the excited states of the system. For the experimentally relevant case of a 3D harmonically trapped gas such that $\kB T/\hbar$ is large compared to the trapping frequencies $\omega_i$ along each axis ($i=x,y,z$), the saturation of the gas is given by \cite{Bagnato:1987}
\begin{equation}
N' \leq \Ncid=\zeta(3)\,\left( \frac{\kB T}{\hbar \bar \omega} \right)^3\ ,
\label{eq:harmonic}
\end{equation}
where $\bar \omega$ is the geometric mean of the $\omega_i$'s and $\zeta$ is the Riemann function [$\zeta(3)\approx 1.202$]. Increasing the total number of particles $\Nt$ above the critical value $\Ncid$ results in the macroscopic occupation of the ground state, i.e. Bose--Einstein condensation.  

The condensation observed in weakly interacting atomic Bose gases is generally believed to provide a faithful illustration of the statistical phase transition proposed by Einstein. At the same time, differences from ideal gas condensation are also observed \cite{ensh96,Gerbier:2004,Gerbier:2004c,Meppelink:2010}, for example in the small deviations of the measured critical atom number $N_c$ from the ideal gas prediction $\Ncid$ \cite{ensh96,Gerbier:2004}. Here we focus on the concept of saturation as the underlying mechanism driving the transition. One might expect that the saturation inequality (\ref{eq:harmonic}) is essentially satisfied in these systems, with just the value of the bound on the right-hand side slightly modified. We prove that this is far from being the case, and show how to reconcile experimental findings with the prediction (\ref{eq:harmonic}). The crucial step in our work is a proper  disentanglement of the subtle role of interactions in condensation. While it is correct that Einstein's statistical argument does not explicitly invoke interactions between the particles, it does assume that the gas is in thermal equilibrium, which is fundamentally impossible to attain in a completely non-interacting system \cite{photongas}.  We overcome this problem by making measurements at a range of interaction strengths, always sufficient to ensure thermal equilibrium, and then extrapolating our results to the non-interacting limit.

\begin{figure}[tbhp]
\begin{center}
\vspace{-7mm}
\includegraphics[width=\columnwidth]{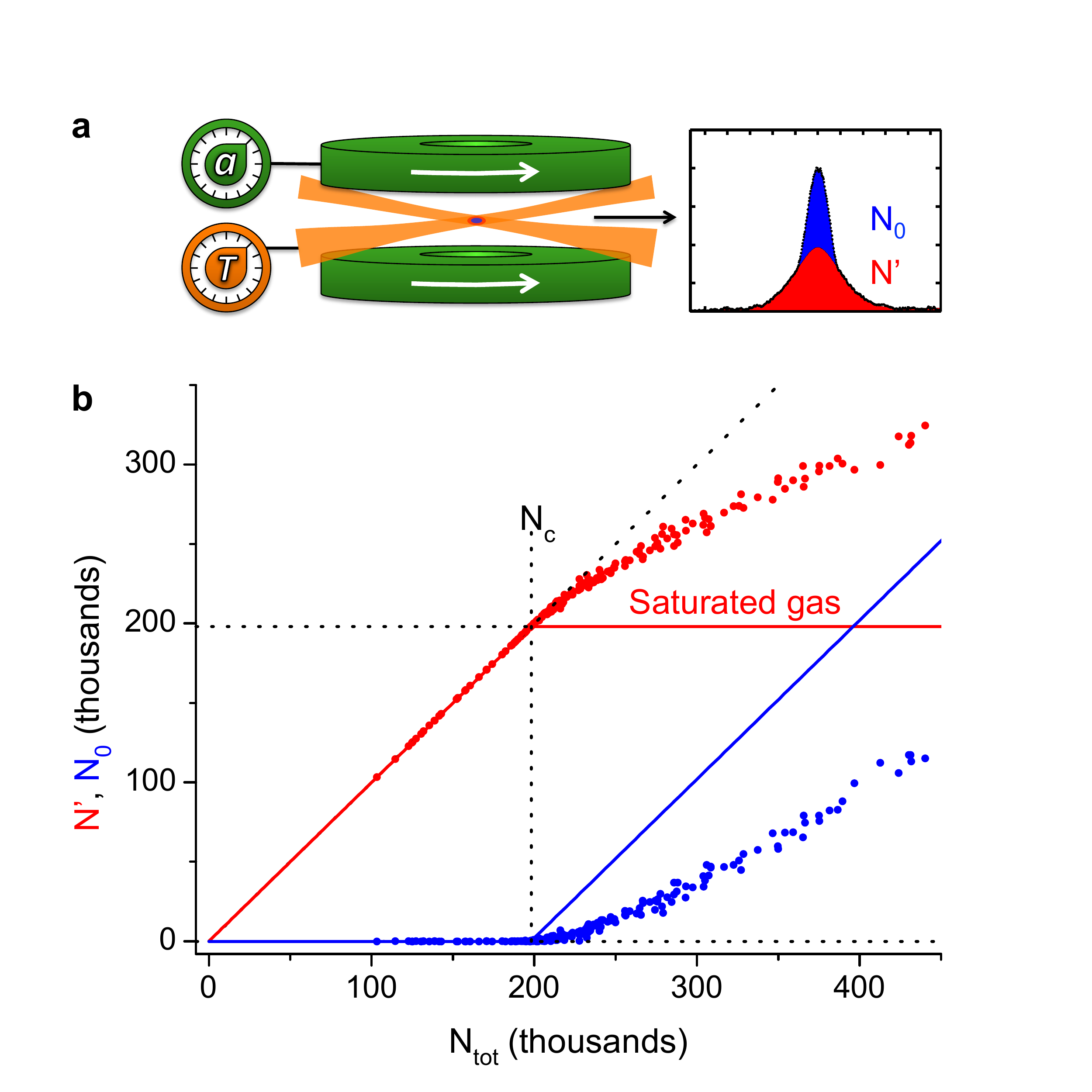}
\end{center}
\vspace{-5mm}
\caption{
\textbf{Lack of saturation of a quantum degenerate atomic Bose gas.}
(a) Schematic of the experimental setup. Potassium ($^{39}$K) 
atoms are confined in an optical dipole trap formed by intersecting two infrared laser beams. The temperature of the gas is fixed by the depth of the trapping potential, and the strength of interactions is controlled by a uniform magnetic field produced by a pair of coils in  Helmholtz configuration. The number of thermal ($N'$) and condensed ($N_0$) atoms is extracted from the density distribution after free expansion from the trap (see Methods). 
(b) $N'$ (red points) and $N_0$ (blue points) versus the total atom number $\Nt$ 
at a temperature $T=177$\,nK and a scattering length $a=135\,a_0$. The corresponding theoretical predictions for a saturated gas are shown by red and blue solid lines. The critical point $\Nt=\Nc$ is marked by a vertical dashed line.}
\vspace{-5mm}
\label{fig:figure1}
\end{figure}

We perform conceptually simple experiments in which we keep the temperature of the gas constant and vary the atom number. Our experimental scheme is outlined in Fig.~\ref{fig:figure1}a (see also Supplementary Information). We start with a partially condensed gas of $^{39}$K atoms in the $|F, m_F\rangle = |1,1\rangle$ lower hyperfine ground state, produced in a crossed optical dipole trap 
\cite{Campbell:2010}. The optical potential near the bottom of the trap is close to harmonic, with $\bar \omega/2\pi$ varying between 60\,Hz and 80\,Hz for data taken at different temperatures. 
We tune the strength of repulsive interactions in the gas, characterised by the positive s-wave scattering length $a$, by applying a uniform external magnetic field in the vicinity of a Feshbach scattering resonance centred at 402.5 gauss \cite{Zaccanti:2009}. 
We always prepare the condensed gas at $a= 135\,a_0$, where $a_0$ is the Bohr radius, and then adjust the scattering length to the desired value by changing the applied magnetic field. In a given experimental series, the temperature is kept constant by fixing the depth of our optical trap, and the atom number is varied by holding the gas in the trap for a variable time up to several tens of seconds. During this time the total atom number slowly decays due to three-body recombination, scattering of photons from the trapping laser beams and collisions with the background gas in the vacuum chamber, while elastic collisions among the trapped atoms ensure equilibrium redistribution of particles between the condensate and the thermal gas.

An example of an experimental series, taken at $a=135\,a_0$ and $T=177$\,nK, is shown in Fig.~\ref{fig:figure1}b. For each hold time between 1 and 110 seconds we extract the number of atoms in the condensate, $N_0$, and in the thermal gas, $N'$, from a bimodal fit to the density distribution of the gas after 18 ms of free time-of-flight expansion from the optical trap. We plot $N_0$ and $N'$ versus the total number of atoms, $N_{\rm tot}$, which is extracted independently by a direct summation over the density distribution. We find that the relationship $N_{\rm tot} = N_0 + N'$ is satisfied for all data points to within 0.5\%. The standard deviation of the temperature for all the points where the condensate is present is 3\,nK (see Supplementary Information). 

The predictions for the number of condensed and thermal atoms in a saturated gas are shown in Fig.~\ref{fig:figure1}b by the blue and red solid lines, respectively. Specifically, for $\Nt > \Nc$, the thermal atom number $N'$ remains constant and equal to $\Nc$. The deviation of the experimental data from this prediction is striking.  As the total number of atoms is increased from the measured critical value, $\Nc \approx 200,\!000$, to $450,\!000$, only half of the additional atoms accumulate in the condensate. 

\begin{figure}[tbhp]
\begin{center}
\vspace{-5mm}
\includegraphics[width=\columnwidth]{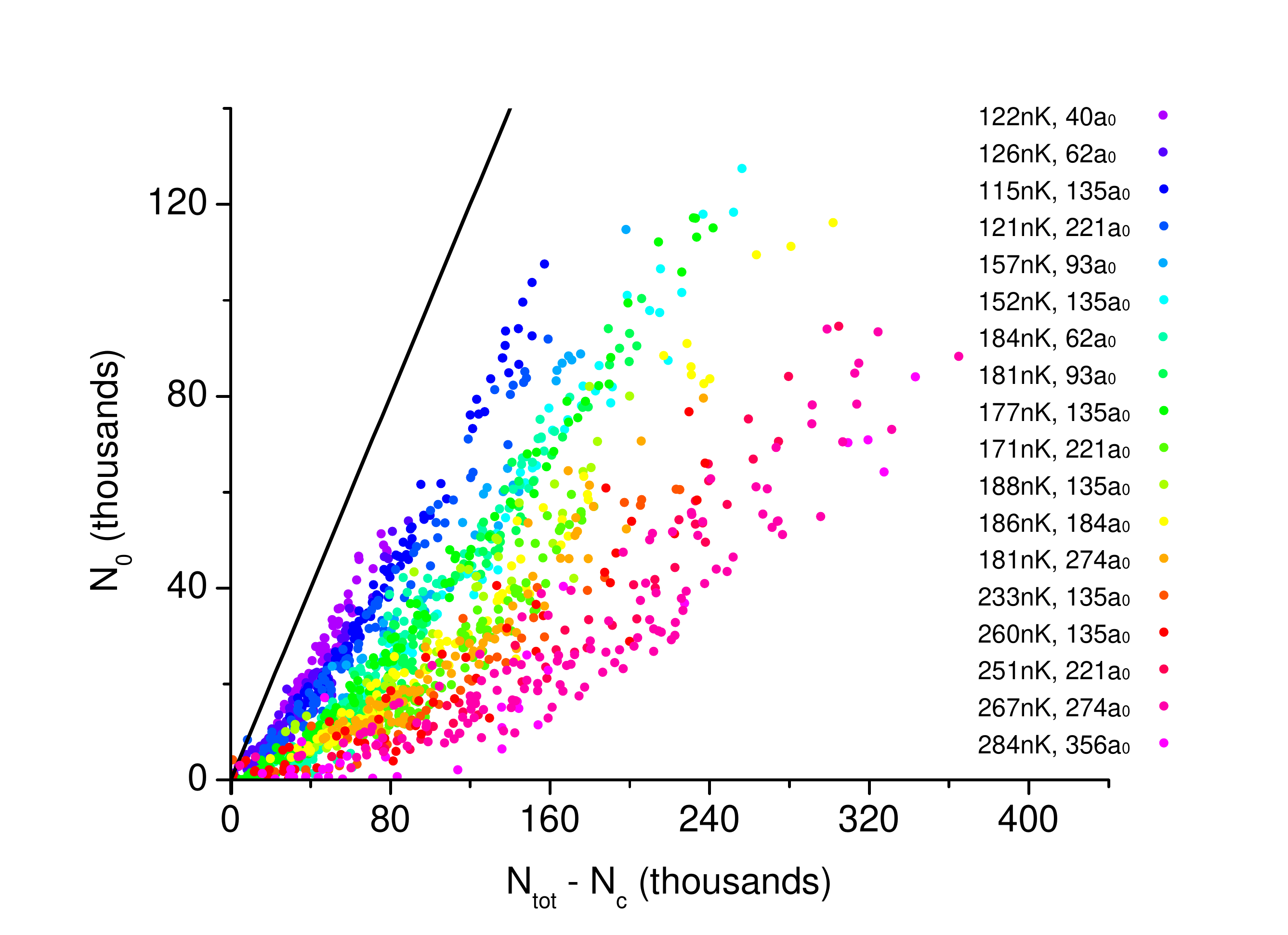}
\end{center}
\vspace{-5mm}
\caption{
\textbf{Deviation from the saturation picture at a range of interaction strengths and temperatures.} 
We plot $N_0$ versus $\Nt-\Nc$ for 18 experimental series, each at fixed $a$ and $T$.
The values of the scattering length (40--356\,$a_0$) and the temperature (115--284\,nK) are encoded in the colour of the data points. The solid line is the prediction for a saturated gas, $N_0=\Nt-\Nc$. }
\vspace{-5mm}
\label{fig:nonU}
\end{figure}

In Fig.~\ref{fig:nonU} we show the results of 18 experimental series 
taken at a wide range of scattering lengths, $40\,a_0<a<356\,a_0$, and temperatures, 115\,nK $<T<$ 284\,nK. 
Here we focus on the regime $\Nt > \Nc$, where the condensate is present, and plot $N_0$ versus $\Nt-\Nc$. The solid line shows the prediction for a fully saturated thermal gas, $N_0=\Nt-\Nc$. The deviation of the data from this prediction is clearly observable in all the series, and grows with both $a$ and $T$.

To explore the relationship between the non-saturation of our Bose gases and the interatomic interactions, we start by identifying the relevant interaction energy. Due to 
the large ratio between the average densities of the condensed and thermal fractions, the non-ideal behaviour of the thermal component primarily results from its interaction with the condensate. The relevant energy scale is then provided by \cite{Dalfovo:1999}:
\begin{equation}
\mu_0=\frac{\hbar \bar \omega}{2}
\left(
15 N_0 \frac{a}{\aho}
\right)^{2/5}\ ,
\label{eq:mu}
\end{equation} 
where $\aho=(\hbar/m\bar \omega)^{1/2}$ is the spatial extension of the ground state of the harmonic oscillator of frequency $\bar \omega$, and $m$ is the atomic mass. The energy $\mu_0$ is the mean-field prediction for the chemical potential of a gas with $N_0$ atoms at zero temperature and in the Thomas-Fermi limit  \cite{Dalfovo:1999}.  
One can understand its scaling with $N_0$ in the following way: For a harmonically trapped condensate the potential energy per particle scales as $E_{\rm P} \propto R^2$, where $R$ is the characteristic size of the cloud, and the mean-field interaction energy per particle is proportional to the density, $E_{\rm I} \propto N_0/R^3$. Minimising the sum of these two energies we get $R \propto N_0^{1/5}$ and $\mu_0 \propto N_0^{2/5}$.

\begin{figure}[tbhp]
\begin{center}
\vspace{-5mm}
\includegraphics[width=\columnwidth]{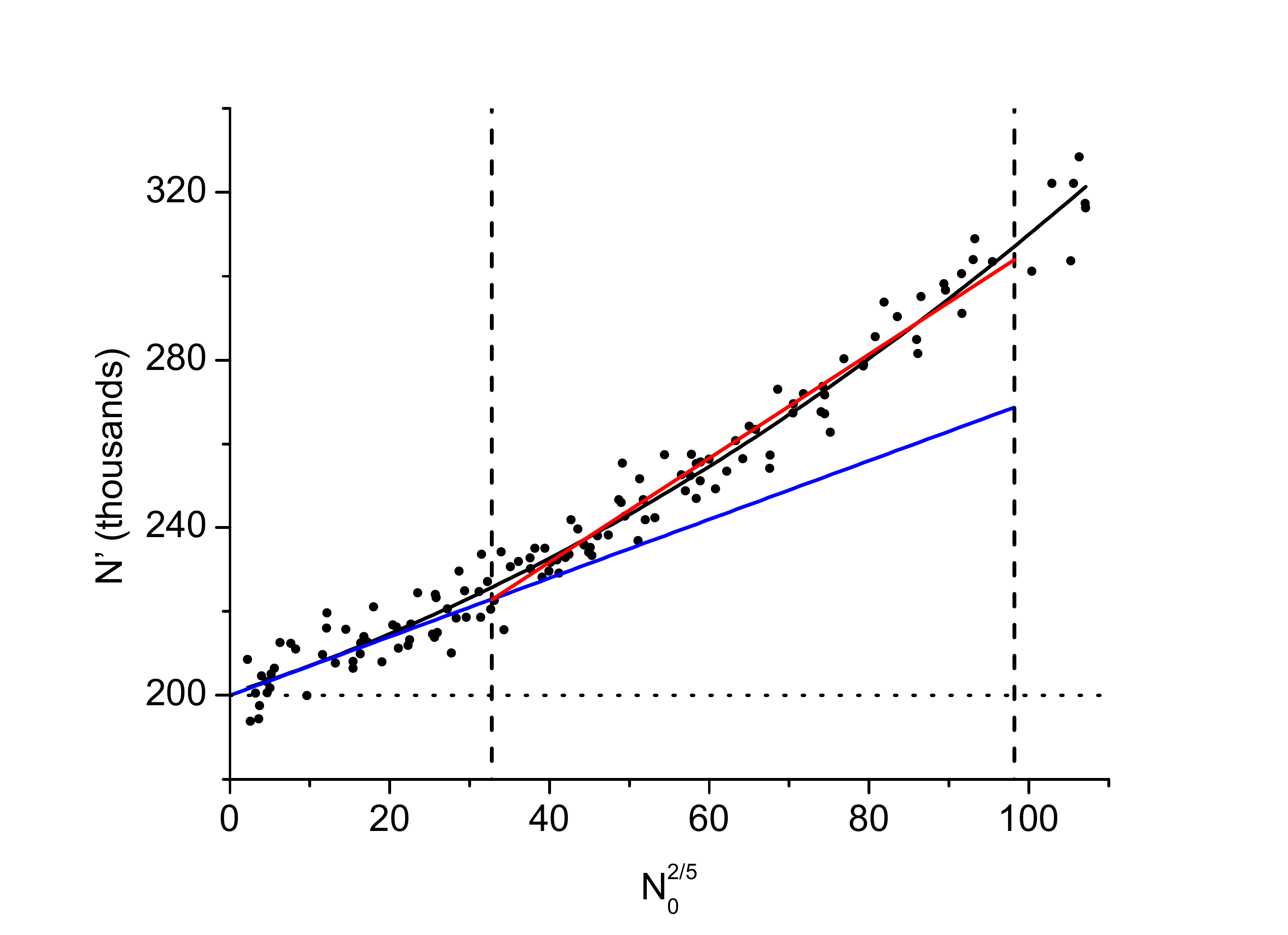}
\vspace{-5mm}
\caption{\textbf{Quantifying the lack of saturation.} Here $N'$ is plotted as a function of $N_0^{2/5}$ for the same series as in Fig.~\ref{fig:figure1}b. The horizontal dotted line is the saturation prediction $N'=\Nc$. The blue line is the mean-field Hartree--Fock result for a harmonically trapped gas (see text), with a slope $S_{\rm HF} = 699$. The red line is a linear fit to the data in the range corresponding to $0.1 < \mu_0/\kB T< 0.3$, which gives a non-saturation slope $S = 1283 \pm 84 $. The solid black line is a guide to the eye based on a second-order polynomial fit. The initial slope of this line is indistinguishable from HF theory.}
\vspace{-5mm}
\label{fig:TwoFifths}
\end{center}
\end{figure}

Guided by this scaling,
in Fig.~\ref{fig:TwoFifths} we plot the thermal atom number $N'$ as a function of $N_0^{2/5}$, for the same experimental series as shown in Fig.~\ref{fig:figure1}b. 
From here we proceed in two steps: First, we show that the initial linear increase of $N'$ with $N_0^{2/5}$ can be quantitatively accounted for by the mean-field Hartree-Fock (HF) theory for a harmonically trapped gas. Second, for the regime of larger condensates, where the theory does not fully reproduce the experimental data, we adopt a more heuristic approach that still allows us to prove the concept of a saturated gas in the non-interacting limit.

In the HF approach one treats the thermal fraction as an ideal gas, but takes into account repulsive interactions with the condensate. Within this theory \cite{Giorgini:1997b,Dalfovo:1999} one gets $N_c = \Ncid$, and can predict a linear variation of $N'/N_c$ with the small parameter $\mu_0\, /\,\kB T$:
\begin{equation}
\frac{N'}{N_c}=1 + \alpha  \, \frac{\mu_0}{\kB T} \; ,
\label{eq:HF}
\end{equation}
with $\alpha = \zeta(2)/\zeta(3)\approx 1.37$ (see Supplementary Information). 
The origin of this non-saturation can be qualitatively understood by noting that interactions with the condensate modify the effective potential seen by the thermal atoms from a parabola into a ``Mexican hat" shape; this allows the thermal component to occupy a larger volume, which grows with increasing $N_0$ \cite{sengstock}.

From Eqs.~(\ref{eq:mu}) and (\ref{eq:HF}) we define the non-saturation slope $S_{\rm HF} = dN'/d(N_0^{2/5}) = 1.37 X$, where $X$ is a dimensionless parameter of the form $X=\xi\, T^2 a^{2/5}$, with $\xi=0.5\;\zeta(3)\, 15^{2/5}\, (\kB/\,\hbar\bar \omega)^{2} \aho^{-2/5}$. The blue line in Fig.~\ref{fig:TwoFifths} corresponds to this prediction, with the intercept fixed by the measured $\Nc$. It agrees with the data very well for small condensates, with $N_0 \lesssim 10^4$, corresponding to $\mu_0\, /\,\kB T \lesssim 0.1$. 

\begin{figure}[h]
\begin{center}
\vspace{-1mm}
\includegraphics[width=\columnwidth]{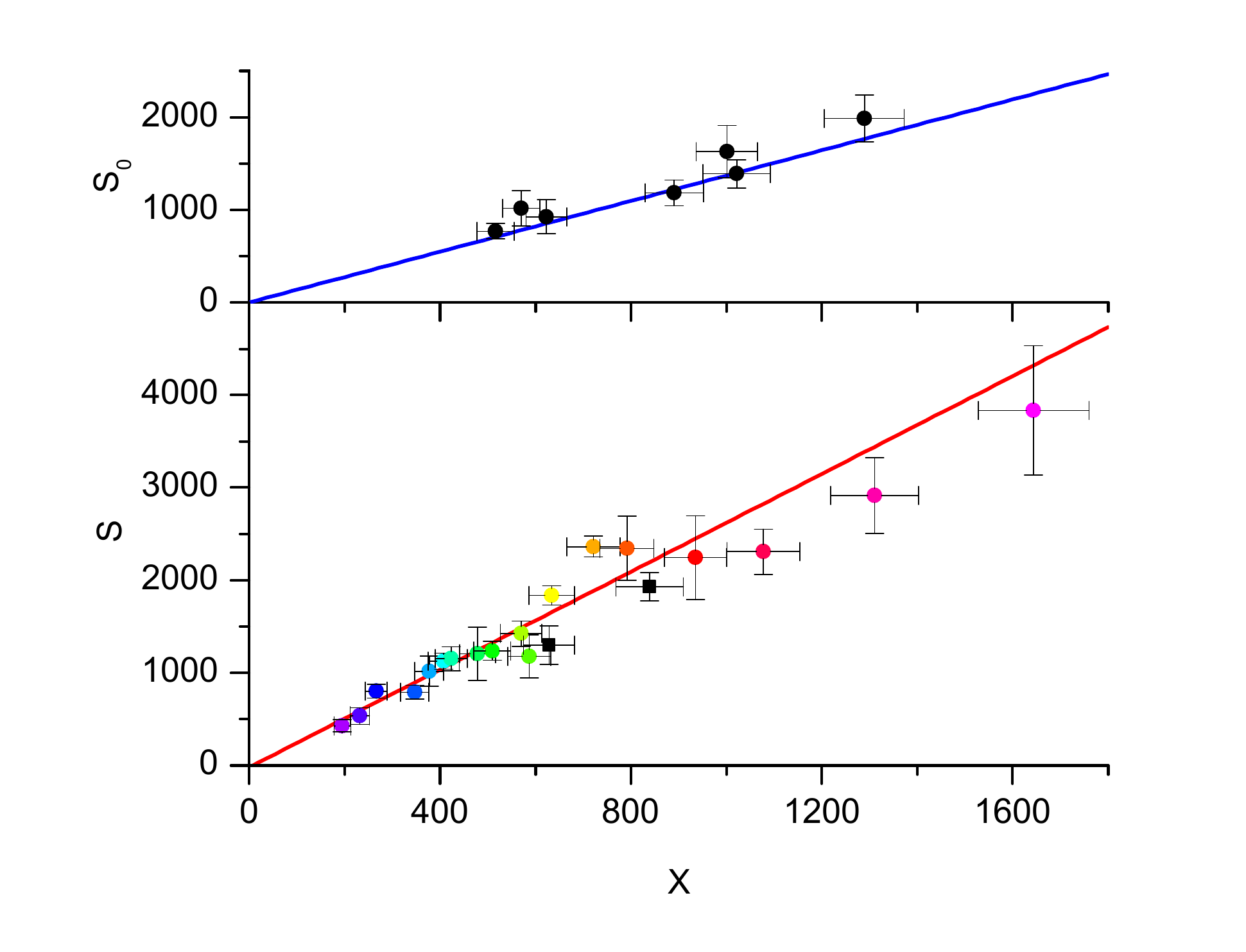}
\vspace{-8mm}
\caption{\textbf{Saturation in the non-interacting limit.} The non-saturation slopes $S_0$ 
and $S$ 
are plotted versus the dimensionless parameter $X=\xi\, T^2 a^{2/5}$ (see text).  The $S_0$ data is directly compared with Hartree-Fock theory,  $S_{\rm HF} = 1.37 X$ (blue line), with no free parameters. For the $S$ data, a linear fit (red line) gives $dS/dX = 2.6 \pm 0.3$\ and an intercept $S(0) = -20 \pm 100$, consistent with complete saturation in the ideal gas limit. The $S$ data is based on the eighteen $^{39}$K series shown with the same symbol code in Fig.~\ref{fig:nonU}, and two additional series taken with $^{87}$Rb (black squares). All vertical error bars are statistical. The systematic uncertainty in atom numbers $N'$ and $N_0$ is $<10\%$, corresponding to $<6\%$ uncertainty in $S_0$ and $S$ values. The horizontal error bars include the 3 Hz uncertainty in the trapping frequencies $\bar \omega/2\pi$ and (for potassium) the 0.1 gauss uncertainty in the position of the Feshbach resonance.}
\vspace{-5mm}
\label{fig:extrapolation}
\end{center}
\end{figure}

To quantitatively test the prediction of Eq.~(\ref{eq:HF}) we took several series at different scattering lengths ($a = 56 - 274 \, a_0$) and temperatures ($T= 177 - 317$ nK), specifically focusing on very small values of $N_0$. We turn off interactions during time-of-flight, so that the small condensates almost do not expand and can be reliably detected and characterised in absorption imaging. For each series we fit the initial non-saturation slope, $S_0 = dN'/d(N_0^{2/5})$ for $N_0 \rightarrow 0$, and compare the result with the prediction $S_{\rm HF} = 1.37 X$. As shown in Fig.~\ref{fig:extrapolation}, the experiment and theory agree within  a few percent.

The agreement of experiments with Eq.~(\ref{eq:HF}) for small $N_0$ is the first main quantitative result of this paper, and allows us to deduce that the initial non-saturation slope $S_0$ would indeed vanish in the non-interacting limit, where $\mu_0 \rightarrow 0$ for any $N_0$. This however does not complete our experimental proof since this theory works very well only for small condensed fractions (see Fig.~\ref{fig:TwoFifths}). For the larger, and experimentally more typical values of $N_0$, we observe an even more pronounced lack of saturation of the thermal gas. 

To quantitatively study non-saturation effects at larger $N_0$ we take the following heuristic approach: Although the observed increase of $N'$ with $N_0^{2/5}$ is not perfectly linear, over a broad experimentally relevant range it can be well quantified by a coarse-grained slope $S= \Delta[N']/\Delta [N_0^{2/5}]$, as indicated in Fig.~\ref{fig:TwoFifths} by the red solid line. In order to treat all the data taken at various values of $a$ and $T$ equally, for each experimental series we consider the same range of values of $\mu_0\, /\,\kB T$, from 0.1 to 0.3. Note that this range covers more than an order of magnitude of $N_0$ values, and encompasses the bulk of the data shown in Fig.~\ref{fig:nonU}.

In Fig.~\ref{fig:extrapolation} we summarise the non-saturation slopes $S(a,T)$ for the same 18 experimental series shown in Fig.~\ref{fig:nonU}. Within experimental error all data points fall onto a straight line when plotted against the dimensionless $X=\xi\, T^2 a^{2/5}$, supporting the assumption that we can still use $\mu_0/\kB T$ as the relevant interaction parameter. 
To further validate our approach, we took additional data 
with a different atomic species, $^{87}$Rb in the $|F, m_F\rangle = |2,2\rangle$ state. In this case $a=99\,a_0$ is not tuneable, and the two experimental series were taken at temperatures of 175\,nK and 203\,nK. These two points are in close agreement with the $^{39}$K data.

The prediction (\ref{eq:harmonic}) for an ideal gas ($a=0$) corresponds to $S=0$ at the origin of the graph in Fig.~\ref{fig:extrapolation}.
The mathematical limit $X \to 0$ is reached in two physically very different limits, $a\to 0$ and $T\to 0$. In the mundane $T \to 0$ limit $N'$  inevitably vanishes for any value of $N_0$, so trivially $S=0$. It is therefore essential that our experiments show that $S$ depends only on the parameter $X\propto T^2 a^{2/5}$, allowing us to deduce its value in the $a \to 0$ limit for any fixed $T$. The solid line in Fig.~\ref{fig:extrapolation} shows the result of an unconstrained linear fit to the data, which gives an intercept consistent with zero, $S(0) = -20 \pm 100$. Together with the success of HF theory in the small $N_0$ regime, this confirms the concept of a saturated Bose gas for a broad range of experimentally relevant parameters. 

We expect our results to be generic to experiments on harmonically trapped 3D Bose gases with (relatively weak)  short-range s-wave interactions. However a number of important questions remain open. Here we studied only the global properties of the gas, inferred from the number of atoms in condensed and thermal components; in the future it would be of great interest to also study saturation at the level of local densities of the two components, and effectively measure the equation of state for a bulk system \cite{Ho:2010a}. Local saturation should depend only on the interactions between particles, but could for example be different in systems with very strong or long-range interactions, where the emergence of a roton minimum significantly increases the density of states for low energy excitations. Global saturation additionally depends on the external potential, and from purely geometric arguments we expect the non-saturation effects to grow with the dimensionality of the system. In this respect it would be particularly interesting to study them for atomic gases in disordered potentials \cite{Sanchez-Palencia:2010}, where the possible fractal nature of the fluid shape can lead to a non-integer effective dimensionality. 

In conclusion, our work shows that the purely statistical picture of a saturated Bose gas is not realised in experiments with harmonically trapped atomic vapours.  However, extrapolation of our results to the strictly non-interacting limit allows us to confirm the textbook picture of Bose-Einstein condensation as a purely statistical phase transition in an ideal gas. Ultracold atomic gases offer great experimental flexibility  for further studies of (non-)saturation effects, which may provide a fruitful way of classifying different geometries and interactions in many-body systems. Better understanding, and ultimately control of these effects could also be of practical importance. Maximising the purity of the condensate for a given attainable temperature and atom number is indeed crucial for minimising the decoherence processes in atom interferometry and quantum information processing.


\vspace{3mm}


\noindent \textbf{Methods Summary} 

The bimodal fit function used to determine $N'$, $N_0$ and $T$ is the sum of the Thomas-Fermi profile for the condensate and the dilogarithm function $g_2$ for the thermal component, with the chemical potential set to zero in the argument of $g_2$~\cite{Ketterle:1999b,Gerbier:2004c}. The fit to the whole distribution is improved by: (1) refitting $N'$ by excluding the central region of the image occupied by the condensate, (2) refitting $N_0$ with fixed parameters of the $g_2$ function, and (3) fitting the temperature to the high energy wings of the distribution by excluding the central region of the cloud corresponding to one thermal radius $R_{\rm T}$. 
Small effects of varying the exclusion region between $0.8\,R_{\rm T}$ and $1.2\,R_{\rm T}$ are included in the error bars.

Within a given series, measurements with different hold times are taken in random order. We sometimes observe small ($<10\%$) temperature drifts as a function of $N_0$ or the time through the series. We analytically correct for these drifts using the temperature scaling given by the Hartree-Fock theory. This amounts to plotting $N' (T/T')^3$ versus $N_0^{2/5}(T/T')$, where $T'$ is assigned to each data point 
and $T$ is the nominal temperature for a series, obtained in the $N_0 \to  0$ limit.

\vspace{5mm}

\noindent *These two authors contributed equally.

\noindent \textbf{Acknowledgements} 
We thank P. Kr\"uger and D. Hutchinson for useful discussions, and N. Cooper, F. Gerbier, M. K\"ohl and C. Salomon for helpful comments on the manuscript. 
This work was support by EPSRC (Grants No. EP/G026823/1 and No. EP/I010580/1). R. P. S. acknowledges support from the Newton Trust. J. D. acknowledges hospitality of the Cavendish Lab and a Visiting Fellowship at Trinity College, Cambridge.

\noindent \textbf{Author Information} 
Correspondence and request for materials should be addressed to Z. H. (zh10001@cam.ac.uk).

\onecolumngrid
\newpage

\begin{center}
\textbf{Supplementary Information}
\end{center}

\vspace{2mm}

\noindent \textbf{Experimental Procedure}

\begin{figure}[h!]
\vspace{-4mm}
\includegraphics[width=0.93\columnwidth]{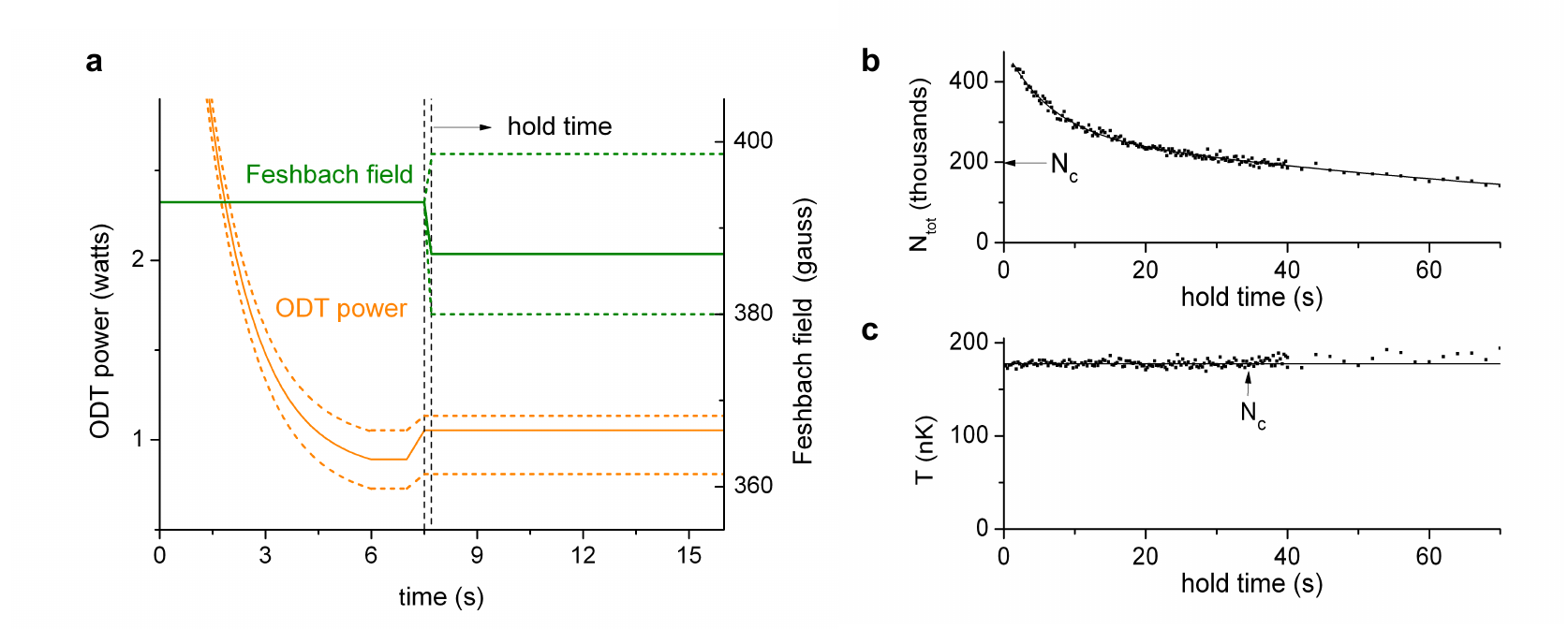}
\vspace{-4mm}
\caption{\textbf{Experimental procedure.} (a) Relevant part of the experimental sequence.  The orange and green lines show the value of the optical dipole trap (ODT) power and Feshbach field respectively as a function of time. The atoms are cooled by an exponential reduction of the ODT power. The power is then ramped up slightly over 0.5 s in order to maintain a steady temperature during the hold time. During the cooling the Feshbach field is held constant to give a scattering length of $135\,a_0$. Once the final ODT power is reached the Feshbach field is ramped to its final value. The range of ODT powers and Feshbach field values used for different experimental series is shown by the dashed lines. (b) Total atom number $\Nt$ versus hold time for the series displayed in the main paper Figs. 1 and 3. The data is fitted with a double exponential to guide the eye. (c) Temperature versus hold time for the same series. The temperature for $\Nt > \Nc$ shows no drift and has a standard deviation of 3nK.}
\end{figure}

\noindent \textbf{Hartree-Fock Theory}


The simplest theoretical framework to address the effects of atomic interactions in a Bose-condensed gas is the Hartree-Fock approximation \cite{Dalfovo:1999SI}. In this model one treats the thermal atoms as a ``non-interacting" gas of density $n'(\bf r)$ confined in the self-consistent mean-field potential  $2g(n_0({\bf r})+n'({\bf r}))$, where $g=4 \pi \hbar^2 a /m$ and $n_0(\bf r)$ is the condensate density. Two further approximations take advantage of the small density of the thermal component: (i) One neglects the influence of the thermal atoms on the condensate spatial distribution, which is then given by $n_0({\bf r}) =\mbox{Max}[(\mu_0-V({\mathbf r}))/g, 0]$ where $\mu_0$ is defined in Eq. (2) of the main paper. This is valid as long as $N_0 \gg \aho/a$ (i.e. $N_0 \gg 100$ to $1000$ for our parameters). (ii) One neglects the mean-field energy $2g n'({\bf r})$ of the thermal atoms, so that the confining potential for these atoms is simply a 3D ``Mexican hat" potential given by
\begin{equation}
V_{\rm eff}({\bf r})=V({\bf r})+2gn_0({\bf r})=|V({\bf r})-\mu_0|+\mu_0\ .
\label{}
\end{equation}
The latter approximation amounts to neglecting the shift of the critical temperature due to interactions.

The number of thermal atoms in the presence of a condensate with $N_0$ atoms 
can now be estimated using the semi-classical version of the Bose--Einstein distribution:
\begin{equation}
N'=\frac{1}{(2\pi \hbar)^3}\int \left\{  \exp\left[\frac{1}{\kB T}\left(\frac{p^2}{2m}+V_{\rm eff}({\bf r})-\mu_0\right)\right]-1 \right\}^{-1}\;
 d^3 r\; d^3p\; .
\label{}
\end{equation}
The result at first order in $\mu_0/\kB T$ reads  (\cite{Dalfovo:1999SI}, Eqs. (120) and (122)):
\begin{equation}
\frac{N'}{N}\approx t^3+\frac{\zeta(2)}{\zeta(3)}t^2 \frac{\mu_0}{\kB \Tcid}\ ,
\label{}
\end{equation}
where $t=T/\Tcid$, and $\kB\Tcid=\hbar \omega (N/\zeta(3))^{1/3}$ is the ideal-gas result for the condensation temperature of a Bose gas with $N$ atoms.  Using $N/\Ncid=t^{-3}$, we can recast the above equation into
\begin{equation}
\frac{N'}{\Ncid}\approx 1+\frac{\zeta(2)}{\zeta(3)} \frac{\mu_0}{\kB T}\ .
\label{}
\end{equation}

\end{document}